# First Experimental Demonstration of Secure NFV Orchestration over an SDN-Controlled Optical Network with Time-Shared Quantum Key Distribution Resources

A. Aguado[1], E. Hugues-Salas[1], P. A. Haigh[1], J. Marhuenda[1], A. B. Price[2,3], P. Sibson[2], J. Kennard[2], C. Erven[2], J. G. Rarity[2], M. G. Thompson[2], A. Lord[4], R. Nejabati[1] and D. Simeonidou[1]

*(1) High Performance Networks group, (2) Centre for Quantum Photonics & (3) Quantum Engineering Centre for Doctoral Training, School of Physics & Department of Electrical and Electronic Engineering, University of Bristol, BS8 1UB ({a.aguado; alasdair.price}@bristol.ac.uk)*
*(4) Head of Optical Research, BT, UK.*

**Abstract** We demonstrate, for the first time, a secure optical network architecture that combines NFV orchestration and SDN control with quantum key distribution (QKD) technology. A novel time-shared QKD network design is presented as a cost-effective solution for practical networks.

**Introduction**

Network function virtualization (NFV) promises significant network infrastructure simplification as current hardware appliances are replaced with software running on standard servers. NFV is complemented by software-defined networking (SDN), provisioning the required network connectivity to respond to newly instantiated appliances by aligning network topologies in an automated manner. However, there are security risks associated with NFV deployment. In an NFV enabled network infrastructure, network functions are stored centrally as software images in a remote data center (DC) where they can be cloned, transferred and deployed as virtual functions on commodity servers (replacing network appliances) across the network. This transfer of network functions must be secured, as any attempt to tamper with NFV can create a significant security breach. For instance, if a transmitted software image of a network function contains any sensitive information, such as a firewall, its interception and/or alteration can compromise an entire network.

Quantum key distribution (QKD) is a contemporary approach to the generation of symmetric keys [1]. In QKD, keys are distributed by transmitting single photons from a sender (Alice) to a receiver (Bob) over a quantum channel, which is usually a fibre optic channel. Fundamental laws of physics prevent an eavesdropper (Eve) from learning the key, as any attempts Eve makes to gain information about the photons will irreversibly change them in a manner that can be readily detected. An additional benefit of QKD is that keys generated in this manner can be considered future-proof from hacking, since they are random rendering any future mathematical attacks ineffective.

Building on these principles, we propose and experimentally demonstrate, for the first time, the inclusion of QKD to tackle NFV's security problems. More importantly, utilizing SDN technology, we present a cost-efficient method for time-sharing the QKD systems and demonstrate the ease in which these systems can be integrated with an NFV platform.

**NFV-QKD platform description**

The ETSI NFV management and orchestration (MANO) architecture [2] is organized into three layers: the orchestrator, virtual network function (VNF) managers, and the virtual infrastructure manager (VIM). Existing solutions approach the functional distribution in several different ways or are not completed, due to a lack of standardization. Here, we define and implement a prototype of the ETSI NFV architecture for distributed DCs (Fig1(a)). Our approach splits the architecture into a centralized orchestrator (CO) acting in master node and different ETSI NFV stacks operating in slave mode. The CO is composed of a Python backend core, mySQL database, a GUI and RESTful interfaces that allow platform users and administrators to manage the infrastructure. The slave-mode stack is composed of three layers: the orchestrator (acting as a gateway between the CO and the DC hosting network functions), the VNF managers, and the VIM, which manages virtual machines (VMs) and Linux containers (based on OpenStack and Docker, respectively).

In the proposed NFV architecture, images of network functions are stored within the CO's trusted DC where an Alice sender unit is also located. To virtualize these network functions, their images must be cloned and securely transmitted to remote servers with Bob receiver units. To achieve secure transmission, our NFV MANO architecture has been integrated with an ID Quantique QKD system (ID3100 Clavis[2]) [3], as shown in Fig.1a. When an image needs to be transmitted, the CO asks for a key from the Alice using the proprietary IDQ3P protocol, before the advanced encryption standard (AES) symmetric key algorithm encrypts the image, informing the remote platform of the transmission and the key ID needed for image decryption. While we use the standard AES algorithm, one-time pad (OTP) could easily be used for mission critical transmissions. The workflow is shown in Fig1(c).

**Time-shared QKD network based on SDN**

As shown in Fig1(b), our novel QKD enabled NFV architecture is proposed in an SDN controlled optical network. This enables us to implement a network control mechanism supporting QKD resource sharing. We implemented a new QKD resource scheduling method, enabling a single physical Alice to be time-shared between multiple endpoints (Bobs) allowing the establishment of multiple secure connections using fewer QKD devices than would normally be required. This method reduces the cost of the proposed secure NFV architecture because of the significant reduction in hardware, and also allows a design in which the most sophisticated components can be contained within the Alice unit, while deploying a number of low-priced Bob units. Fast scheduling of optical connections is achieved by using SDN, which is also employed to synchronize the QKD units, together with the remote NFV stacks and the NFV orchestrator. This scheduling is undertaken assuming that the QKD channels and the classical data channels are multiplexed in space (i.e. carried over different fibers). The scheduling process can be summarized as follows: *i*) The NFV orchestrator identifies the Bob nodes requiring to share a particular Alice node, *ii*) The scheduling mechanism utilizing information extracted from the network SDN controller creates the required end-to-end connectivity for the Alice/Bob pairs by establishing optical connections between the Alice and every Bob, one connection at a time and sequentially, *iii*) the key servers are started and stand by until the QKD initialization process is finished, *iv*) The NFV orchestrator extracts the generated keys from both QKD units, keeping them with identification by the key ID extracted from the response message.

**Experimental test-bed**

Our time-shared QKD test-bed comprises four nodes, as shown conceptually in Fig1(b). This testbed comprises one Alice device in Node 1, together with a key server which is time shared with three Bob devices (Nodes 2 to 4) and their key servers. In each node, keys can be extracted for use in a number of different protocols, such as AES or OTP. The Alice/Bob nodes are connected via a dynamic, and reconfigurable circuit switched optical network. The network comprises a 25ms-switching, large-port count programmable and SDN-enabled (with OpenFlow extensions) beam-steering optical switch (Polatis). This is partitioned into multiple smaller interconnected switches emulating a network. The optical network and the switch are used for directing both standard and quantum signals between nodes. The Alice/Bob units, operating at 1552nm, are controlled by separate servers using a software suite designed for automated hardware operation and key distillation. In the test-bed, the QKD and classical data channels are always running in separated fibers. This test-bed is equipped with three servers running the proposed NFV architecture stack, the SDN controller for the optical network and the QKD device scheduling mechanism.

**Experimental results**

The results from our time-shared QKD network are shown in Fig. 2. Fig 2(a) shows the measured average initialization time for each Alice/Bob pairs, i.e. the time taken for synchronization, characterization of the setup and generation of the first set of keys. The measured average initialization time increases with distance between Alice/Bob pairs in a double exponential function, as shown by the curve fitted ($R^2$ = 0.9999). For the back-to-back (BtB) case, the average initialization time is 400 s, increasing to 1265 s at 25 km, an increase of around threefold, indicating that as the distance (and power penalty) increases, the time taken to initialize becomes substantially larger. This result is particularly useful when time-sharing the Alice unit, as it allows a sensible estimate of the delays encountered when cyclically communicating between different Bob units. For enabling longer distances, a multi-hop approach could be considered using additional trusted nodes. Fig 2(b) shows the secret key rate as a function of distance. The mean values are represented by the circular marker, while the maximum and minimum values are shown as error bars. The secret key rate in the BtB scenario is ~4 kb/s, while in the 25 km case the rate drops to ~0.1 kb/s, due to the additional attenuation. The attenuation itself is shown in Fig. 2(c) on the right hand axis which, as expected, displays an approximately linear response. Fig. 2(c) also shows the quantum bit error rate (QBER), on the left hand side. In line with the other results shown, the longer the distance between Alice/Bob nodes the higher the QBER, affecting the secret key rate with approximately linear proportionality. A QBER of 5.3% is reached for a QKD pair of over 25km of standard single-mode fibre (SSMF).

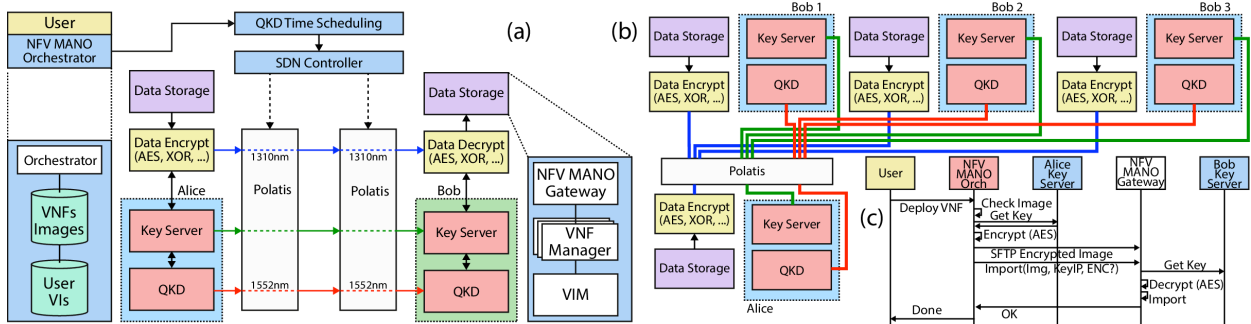

Fig 1: (a) The proposed ETSI NFVI architecture on a QKD enabled optical network test-bed, (b) QKD key extraction, encryption & transmission workflow, (c) Logical representation of the experiment, QKD-Alice is time shared between three QKD-Bobs

Regarding image transmission between the nodes, our setup was composed of three Dell PowerEdge T630 servers (one hosting the orchestrator and scheduling mechanism, one hosting the ETSI NFV stack in slave mode and one hosting the SDN controller), running 1310 nm 10G SPF interfaces, allowing the servers to communicate through our optical network (note any wavelengths can be used here since the data and quantum channels are carried over separate fibres). As shown in Fig 2(e), the transmission of the 16 GB Windows Server image by encrypting (126 s), sending through a standard socket (33 s) and decrypting (144 s) takes ~305 s (which may vary depending on the computational load of the server), while by using secure file transfer protocol transmission takes ~405 s. Fig. 2(d) shows the capture of messages exchanged for the VNF transmission to the remote DC, as shown previously in Fig. 1(c). The first message initiates the workflow, before the orchestrator creates the required network connectivity for the transmission by sending CFlow_mod OpenFlow messages to the Polatis switch through OpenDaylight (highlighted in red). The orchestrator requires the key from the Alice unit to encrypt the image and sends it to the remote DC by using a standard TCP socket (highlighted in green). To finalize the process, the orchestrator informs the image and key IDs to be used for the decryption, sending an ACK (HTTP 200) when the process is finished. Some messages (IDQ3P response, TCP) are omitted to improve readability.

**Conclusions**

In this paper, NFV orchestration over SDN-controlled optical networks with advanced quantum-key-distribution systems is shown for the first time. A particular ETSI-NFV inter-DC architecture is designed on a QKD capable optical network test-bed with a novel SDN-based QKD resource scheduling method for time-sharing a single QKD-Alice between multiple Bob units. Results demonstrate that a SSMF link of up to 25km long can be secured supported by Quantum encryption with simultaneous NFV and SDN control, with a minimum QBER of 5.3%.

**Acknowledgements**

This work acknowledges EPSRC EP/M013472/1: UK Quantum Technology Hub for Quantum Communications Technologies and EP/L020009/1: Towards Ultimate Convergence of All Networks

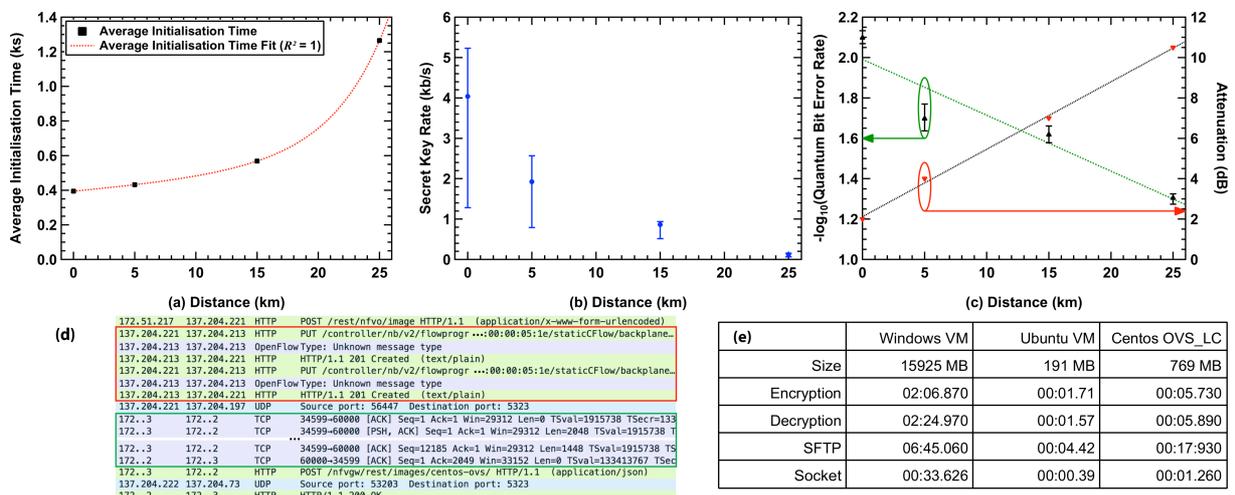

Fig 2 a) time shared Alice node with three Bob nodes, (b) secret key rate vs distance (c) QBER and attenuation vs distance, (d) WireShark capture for the image transmission and (e) times for different transmissions